\documentclass[journal=nalefd,manuscript=article]{achemso}

\usepackage{chemformula} % Formula subscripts using \ch{}
\usepackage[T1]{fontenc} % Use modern font encodings
\usepackage{graphicx}

\author{Fatima Ibrahim}
\affiliation{Univ. Grenoble Alpes, CEA, CNRS, SPINTEC, Grenoble, France}
\email{fatima.ibrahim@cea.fr}
\author{Ali Hallal}
\affiliation{Univ. Grenoble Alpes, CEA, CNRS, SPINTEC, Grenoble, France}
\author{Daniel Solis Lerma}
\affiliation{Univ. Grenoble Alpes, CEA, CNRS, SPINTEC, Grenoble, France}
\author{Xavier Waintal}
\affiliation{Univ. Grenoble Alpes, CEA, IRIG-PHELIQS, Grenoble, France}
\author{Evgeny Y. Tsymbal}
\affiliation{Department of Physics and Astronomy and Nebraska Center for Materials and Nanoscience, University of Nebraska, Lincoln, NE, USA}
\author{Mairbek Chshiev}
\affiliation{Univ. Grenoble Alpes, CEA, CNRS, SPINTEC, 38000 Grenoble, France}
\email{mair.chshiev@cea.fr}
\phone{}
\fax{}

\title[An \textsf{achemso} demo]
  {Unveiling multiferroic proximity effect in graphene}

\begin{document}

%%%%%%%%%%%%%%%%%%%%%%%%%%%%%%%%%%%%%%%%%%%%%%%%%%%%%%%%%%%%%%%%%%%%%
%% The "tocentry" environment can be used to create an entry for the
%% graphical table of contents. It is given here as some journals
%% require that it is printed as part of the abstract page. It will
%% be automatically moved as appropriate.
%%%%%%%%%%%%%%%%%%%%%%%%%%%%%%%%%%%%%%%%%%%%%%%%%%%%%%%%%%%%%%%%%%%%%
%\begin{tocentry}

%Some journals require a graphical entry for the Table of Contents.
%This should be laid out ``print ready'' so that the sizing of the
%text is correct.

%Inside the \texttt{tocentry} environment, the font used is Helvetica
%8\,pt, as required by \emph{Journal of the American Chemical
%Society}.

%The surrounding frame is 9\,cm by 3.5\,cm, which is the maximum
%permitted for  \emph{Journal of the American Chemical Society}
%graphical table of content entries. The box will not resize if the
%content is too big: instead it will overflow the edge of the box.

%This box and the associated title will always be printed on a
%separate page at the end of the document.

%\end{tocentry}

%%%%%%%%%%%%%%%%%%%%%%%%%%%%%%%%%%%%%%%%%%%%%%%%%%%%%%%%%%%%%%%%%%%%%
%% The abstract environment will automatically gobble the contents
%% if an abstract is not used by the target journal.
%%%%%%%%%%%%%%%%%%%%%%%%%%%%%%%%%%%%%%%%%%%%%%%%%%%%%%%%%%%%%%%%%%%%%
\begin{abstract}
We demonstrate that electronic and magnetic properties of graphene can be tuned via proximity of multiferroic substrate. Our first-principles calculations performed both with and without spin-orbit coupling clearly show that by contacting graphene with bismuth ferrite BiFeO$_3$ (BFO) film, the spin-dependent electronic structure of graphene is strongly impacted both by the magnetic order and by electric polarization in the underlying BFO. Based on extracted Hamiltonian parameters obtained from the graphene band structure, we propose a concept of six-resistance device based on exploring multiferroic proximity effect giving rise to significant proximity electro- (PER), magneto- (PMR), and multiferroic (PMER) resistance effects. This finding paves a way towards multiferroic control of magnetic properties in two dimensional materials. 
\end{abstract}

%%%%%%%%%%%%%%%%%%%%%%%%%%%%%%%%%%%%%%%%%%%%%%%%%%%%%%%%%%%%%%%%%%%%%
%% Start the main part of the manuscript here.
%%%%%%%%%%%%%%%%%%%%%%%%%%%%%%%%%%%%%%%%%%%%%%%%%%%%%%%%%%%%%%%%%%%%%
%\section{Introduction}
Spintronic devices possessing high speed and low-power consumption have opened new prospects for information technologies. As the spin generation, manipulation, and detection is the operating keystone of a spintronic device, realizing those three components simultaneously stands as a major challenge limiting applications \cite{Chappert,Zutic,Lou,Dash}. In this context, developing a suitable spin transport channel which retains both long spin lifetime and diffusion length is highly desirable. Graphene stands as a potential spin channel material owing to its exceptional physical properties. Beside its high electron mobility and tunable-charge carrier concentration, graphene has demonstrated room temperature spin transport with long spin-diffusion lengths ~\cite{Geim,Neto,Tombros,Popinciuc,Dlubak1,Han1,TY-Yang,Maasen,Dlubak2,Cummings,Tuan}.  Accordingly, graphene spintronics became a promising direction of innovation that attracted a growing attention in the scientific community ~\cite{Roche,Han2}. 

Much efforts have been devoted to induce magnetism in graphene via different means ~\cite{Yazyev1,Yazyev2,Yazyev3,Son,Kim1,Bai,HX-Yang1,Trolle,Soriano,McCreary,Chan,Ding,Zhang,Jiang,Kim2,JW-Yang}, one of which is the exchange-proximity interaction with magnetic insulators ~\cite{ZWang,Leutenantsmeyer,Singh}. Theoretically, this effect was demonstrated using different materials such as ferromagnetic ~\cite{HX-Yang2,Sakai}, antiferromagnetic ~\cite{Qiao}, topological ~\cite{Vobornik}, and multiferroic ~\cite{Zanolli1} insulators where exchange-splitting band gaps reaching up to 300 meV were demonstrated.  Recently, a detailed study has shown the influence of different magnetic insulators on the magnetic proximity effect induced in graphene ~\cite{Hallal}. On the other hand, experiments on YIG/Gr ~\cite{ZWang,Mendes,Leutenantsmeyer,Evelt}, EuS/Gr ~\cite{Wei}, and BFO/Gr ~\cite{Wu,Song1} demonstrated proximity induced effect in graphene with substantial exchange field reaching $14$ T. However, combining both conditions of a high Curie temperature ($T_c$) magnetic insulator and a weak graphene doping stands as a major challenge which limits practical spintronic applications.                   

Multiferroics, co-exhibiting a magnetic and ferroelectric order, constitute an interesting class of magnetic insulators that bring about an additional parameter in play which is the electric polarization.  On one hand, proximity induced magnetism was reported in graphene using multiferroic magnetic insulator ~\cite{Zanolli1,Zanolli2,Qiao} ignoring the influence of electric polarization. On the other hand, the ferroelectrically-driven manipulation of the carrier density in graphene was demonstrated ~\cite{Baeumer}. However, the ferroelectric control of magnetic proximity effect has not been addressed so far. In this letter, we report the multiferroic-induced proximity effect (MFPE) in graphene proposing the concept of controlling electronic and magnetic properties of graphene via multifferoic substrate. For this purpose, we considered bismuth ferrite BiFeO$_3$ (BFO) whose room-temperature multiferroicity promotes it as a good candidate for applications ~\cite{Neaton,Ravindran,JWang,Zavaliche,Bea}. Our first-principles calculations demonstrate that by contacting graphene with BFO, the spin-dependent electronic structure of graphene is highly influenced not only by the magnetic order but also by the ferroelectric polarization in the underlying BFO. These findings propose additional degrees of control for proximity induced phenomena in graphene and perhaps in other two-dimensional materials.  
   
\begin{figure}[t]
  \centering
     \includegraphics[width=0.9\textwidth]{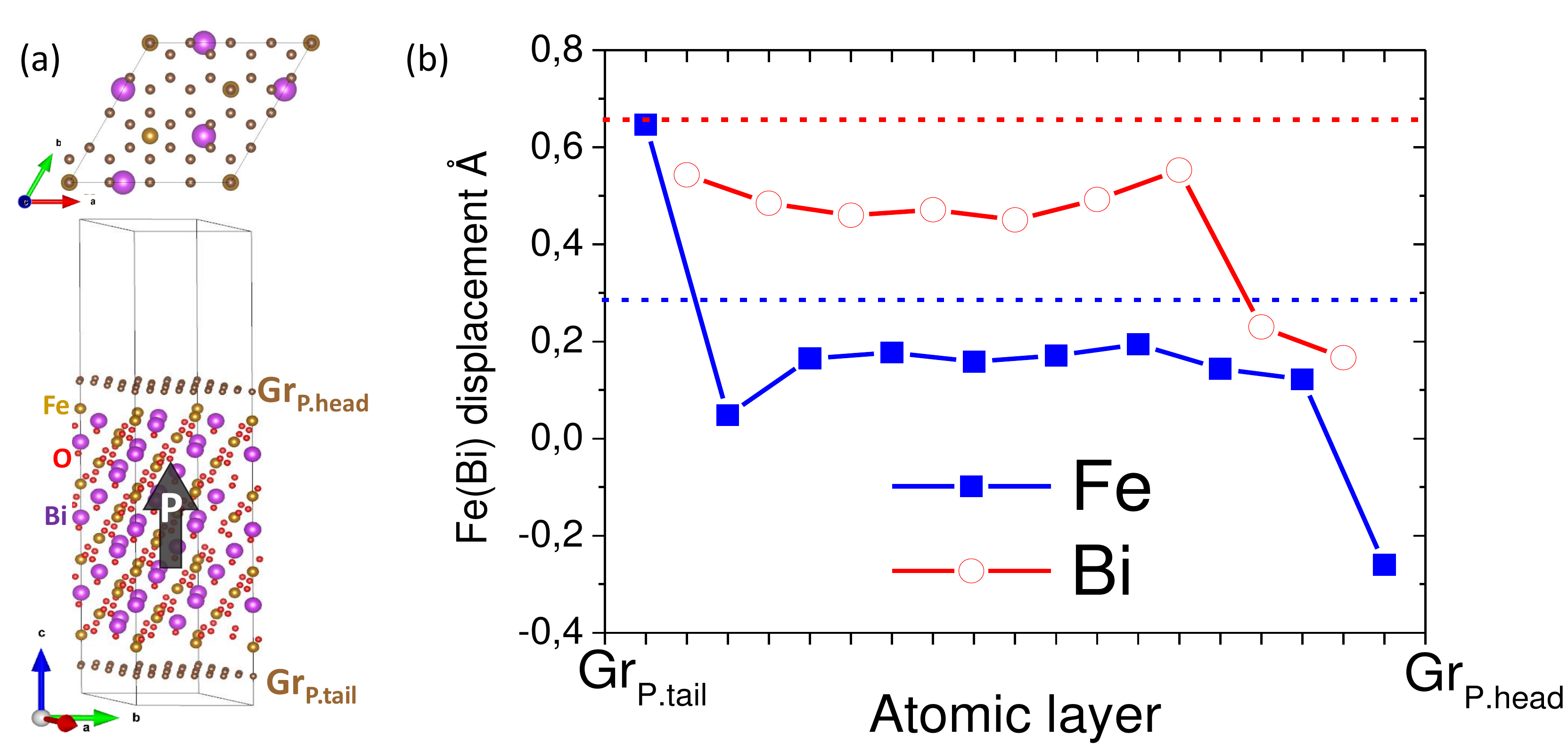}
  \caption{(Color online) (a) The Gr$_{P.head}$/BFO/Gr$_{P.tail}$ supercell is shown in the lower panel. Magenta (Gold) balls designate Bi (Fe) atoms respectively and small red balls represent O atoms. A top view of the Gr/BFO interface is shown in the upper panel where one Fe atom occupies a hollow site and the other two occupy top sites. (b) The layer-by-layer Fe(Bi) displacement from their centrosymmetric positions shown by square (circle) symbols. The blue (red) dashed lines correspond to the bulk values of the displacements for Fe(Bi). The direction of the electric polarization originating from these atomic displacements is perpendicular to the interface, along the $c$-axis, and shown by an arrow. }
  \label{fig1}
\end{figure}
 
Our first-principles calculations are based on the projector-augmented wave (PAW) method ~\cite{Blochl} as implemented in the VASP package ~\cite{Kresse1, Kresse2, Kresse3} using the generalized gradient approximation as parametrized by Perdew,Burke, and Ernzerhof \cite{Perdew,Kresse4}. A kinetic energy cutoff of 550 eV has been used for the plane-wave basis set and a $9\times 9\times 1$ $k$-point mesh to sample the first Brillouin zone. The supercell comprises of nine (Bi-O$_{3}$-Fe) trilayers of BFO ($111$) surface with Fe termination sandwiched between two $4\times 4$ graphene layers as shown in Figure ~\ref{fig1} (a). We fixed the in plane lattice parameter to that of BFO where the lattice mismatch in this supercell configuration is about $1.5\%$. This heterostructure provides the opportunity to compare simultaneously the properties of two different graphene layers relatively sensing opposite directions of the BFO polarization $P$. Since maintaining the polarization is a critical issue in ferroelectric  slabs, a thick BFO slab is used both to restore the electric polarization within the bulk layers and to assure that the two graphene layers do not interact. At both Gr/BFO interfaces, one Fe atom is placed at a hollow site whereas the other two atoms occupy top sites as shown in the top view of Figure~\ref{fig1} (a).   Then, the atoms were allowed to relax in all directions until the forces became lower than $1$ meV/\AA{}. As the GGA fails to describe the electronic structure of strongly correlated oxides, we have employed the GGA+U method to the Fe-$3$d orbitals ~\cite{Liechtenstein}. We have optimized the value of $U$ using the bulk unit cell of BFO and found that $U_{eff}=4$ eV yields $2.44$eV band gap and $\pm 4.15$ $\mu_{B}$/Fe magnetic moments which are in good agreement with experimental values ~\cite{Sosnowska,Kubel,Ihlefeld}.

\begin{figure}[t]
  \centering
     \includegraphics[width=0.9\textwidth]{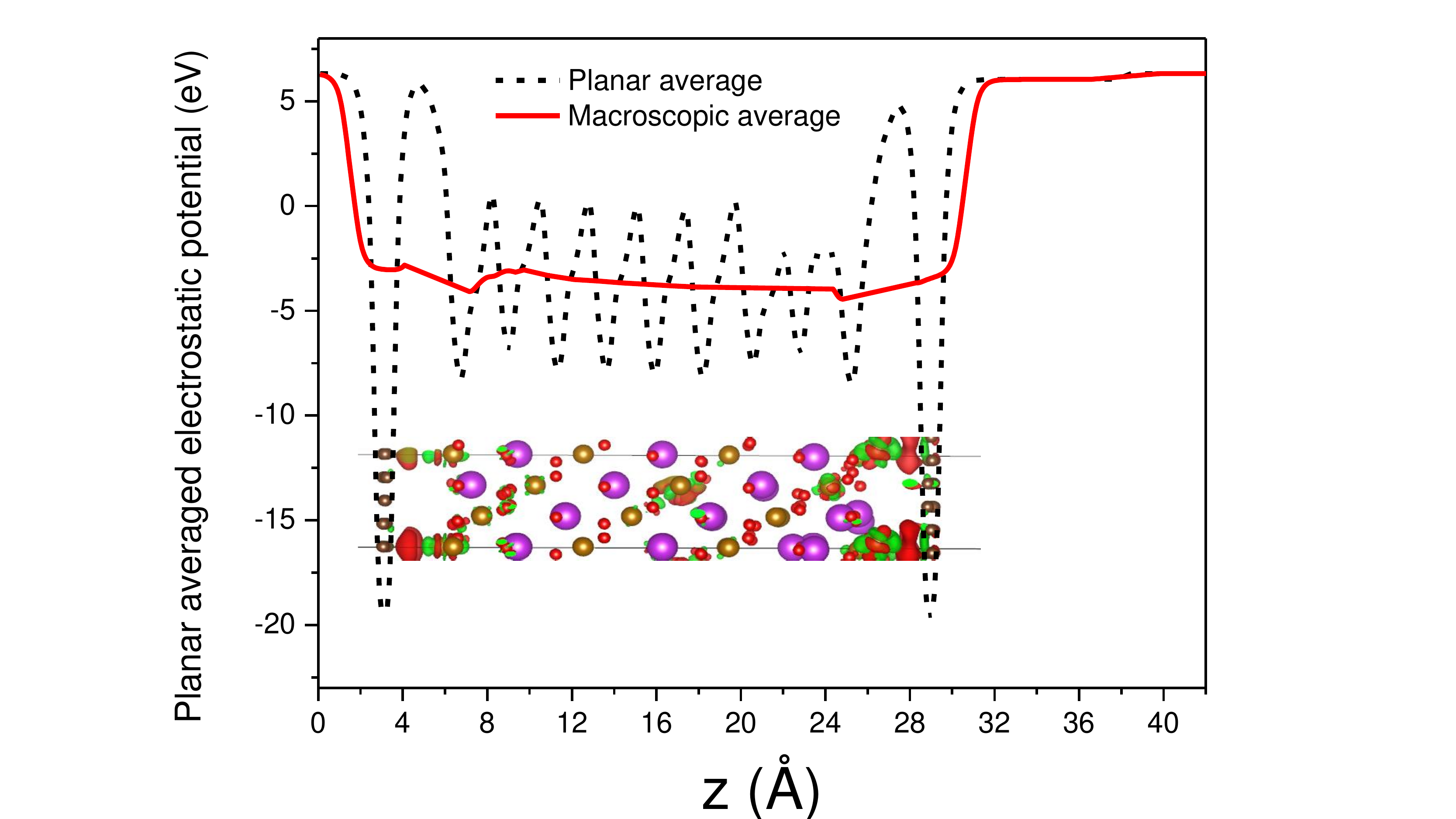}
  \caption{(Color online) The calculated planar averaged electrostatic potential (dashed black line) and its macroscopic average (solid red line) across the Gr/BFO/Gr supercell. The inset shows the induced spatial charge distribution upon the formation of the two Gr/BFO interfaces. The red (green) regions indicate charge accumulation (depletion), respectively. the plot is obtained using an isovalue$=0.002$ e/\AA{}.}
  \label{fig2}
\end{figure}

BiFeO$_{3}$ has a perovskite type cryctal structure and belongs to the polar space group $R3c$. The spontaneous polarization $P$ along BFO (111) direction originates from the displacements of the Bi and Fe atoms from their centrosymmetric positions along the (111) direction ~\cite{Kubel,JWang,Neaton}. To examine $P$ of BFO after the formation of the Gr/BFO/Gr interfaces which accounts for both the ionic and charge relaxation, we show in Figure \ref{fig1}(b) the Fe and Bi $z$-displacements from their centrosymmetric positions per atomic layer. It can be clearly seen that the two BFO/Gr interfaces have different values of atomic displacements whereas in the bulk layers the values are almost constant in good relevance to the bulk values (shown by dashed lines). This infers that $P$, which arises from such non-centrosymmetric structure, is sustained in BFO and it is perpendicular to the interface and pointing from lower graphene layer, lying at the tail of $P$ and denoted hereafter by Gr$_{P.tail}$, towards the upper one lying at the head of $P$ denoted by Gr$_{P.head}$.  A rough estimate of the $z$-averaged polarization can be deduced from the values of the local polarization based on Born effective charges: $P(z)=\frac{e}{\Omega}\sum_{m=1}^{N} Z_{m}^{*} \delta z_{m}$ ; where N is the number of atoms, $\delta z_{m}$ is the displacement of the $m$th atom from the centrosymmetric position, $\Omega$ is the volume of the unit cell, and $Z^{*}_{m}$ is the Born effective charge of the $m$th ion. In our supercell a value of $P=63$ $\mu$C/cm$^2$ is estimated which reasonably compares to the calculated value in a bulk BFO unit cell $P=100$ $\mu$C/cm$^2$ ~\cite{Neaton}.    

\begin{figure*}[th]
  \centering
     \includegraphics[width=1\textwidth]{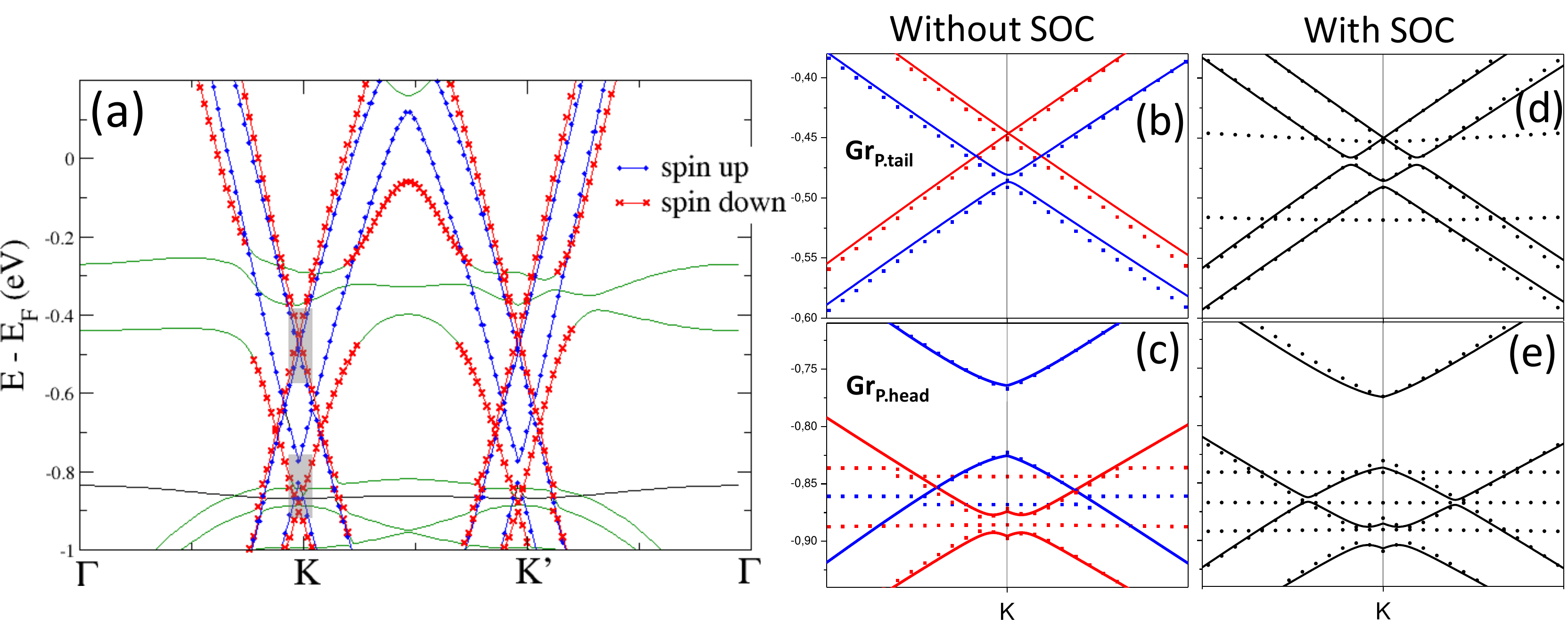}
  \caption{(Color online) (a) Calculated band structure for Gr/BFO/Gr heterostructure without including spin-orbit interactions. Spin up (spin down) bands are shown in blue diamond (red cross) lines, respectively. (b), (c) are zoom around K point shown by the shaded areas in (a) corresponding to the Dirac cones for Gr$_{P.tail}$ and Gr$_{P.head}$, respectively. (d), (e) are the band structure calculated by including spin-orbit coupling shown for the same shaded region as in (b,c) for comparison. The dotted symbols and solid lines in (b, c, d, and e) correspond to the DFT calculated and tight-binding fitted band structures, respectively.} 
  \label{fig3}
\end{figure*}

We discuss now the formation of Gr/BFO/Gr interface. The BFO(111) slab is Fe$^{3+}$ terminated on both sides which makes the two surfaces polar with a nonzero net charge. From a macroscopic electrostatic point of view, this is equivalent to a slab having a polar surface charge $\sigma_{s}= +1.5e/A=88 \mu C/cm^{2}$  on both surfaces and no charges inside the slab, where $A$ is the surface area per Fe atom. On the other hand, assuming a uniform polarization $P$ in the BFO slab whose direction is shown in Figure \ref{fig1}(a) yields surface polarization charges  $\sigma_{P}= +P$ and $\sigma_{P}= -P$  on the {\it head} and {\it tail} surfaces, respectively. Therefore, the whole BFO slab is equivalent to a slab with total bound charge  $\sigma_{head} = \sigma_{s} + P = 151 \mu C/cm^{2}$ on the {\it head} surface and $\sigma_{tail} = \sigma_{s} - P=25 \mu C/cm^{2}$ on the {\it tail} surface. This dissimilarity in the BFO surface charges leads to the formation of two significantly different interfaces with graphene giving rise to two adsorption distances $\Delta z(Gr_{P.head}-BFO)=2.35$\AA{} compared to $\Delta z(Gr_{P.tail}-BFO)=2.7$\AA{}. In fact, graphene sheets adsorbed on both sides of the slab accumulate negative charges trying, ideally, to screen the positive bound charges on the BFO surfaces. This produces a strong electrostatic interaction between graphene and the BFO surfaces in particular at the {\it head} interface where the bound charges are quantitatively larger as shown in Figure \ref{fig2}. Consequently, (i) the $G_{P.head}$ relaxes closer to the BFO surface compared to $G_{P.tail}$ and (ii) strong relaxations are induced at the {\it head} BFO surface revelaed by the smaller polar displacements at the outermost layers, as shown in Figure \ref{fig1}(b), thus, reducing the effective polarization at this surface.  

To get more insights on the interaction at the Gr/BFO/Gr interfaces, the inset of Figure \ref{fig2} shows the induced charge distribution upon the formation of the interfaces. Negative charges, represented by red regions, are accumulated at both Gr/BFO interfaces in accord with the description we provided in the previous paragraph. However, the charges at the Gr$_{P.head}$ are obviously larger than at the Gr$_{P.tail}$. This is a direct implication of the stronger electrostatic interaction at the {\it head} interface which is responsible for the shorter interfacial distance. 

\begin{table*}[th]
\centering
\begin{tabular}{lcccccccccc}
%\begin{tabular}{1\textwidth}{ l c c c c c c c c c c c c c c  }
  \hline
                       &	$E_{G}$ & $\Delta_{\uparrow}$	& $\Delta_{\downarrow}$	& $\delta_{e}$& $\delta_{h}$	& $E_{D}$& $\gamma_{soc}$ & $t_{\uparrow}$ & $t_{\downarrow}$& $t_R$   \\	
\hline
 Gr$_{P.head}$ &	 -48.6 & 55   & 26   &  104   & 75 &  -0.85   &  4   & 2.66 & 2.3 & 8.7 \\
                       &          &        &        &         &      &            &       &2.66 & 2.28& \\
                       &          &        &        &         &      &            &        &2.61 & 2.32& \\         
\hline         
 Gr$_{P.tail}$    & -34.04& 6     & 1.5   & -35  & -40 	&  -0.47   &  5   & 2.42& 2.5& 7.5  \\
 \hline
\end{tabular}
\label{tab1}
\caption{Extracted energy gaps and exchange splitting parameters of Gr$_{P.head}$ and Gr$_{P.tail}$ at Dirac point for Gr/BFO/Gr heterostructure. $E_{G}$ is the band-gap of the Dirac cone given in units of meV. $\Delta_{\uparrow}$ and $\Delta_{\downarrow}$ are the spin-up and spin-down gaps in meV, respectively. The spin-splitting in meV of the electron and hole bands at the Dirac cone are $\delta_{e}$ and $\delta_{h}$, respectively. $E_{D}$ in eV is the Dirac cone position with respect to Fermi level. $\gamma_{soc}$ denotes the spin-orbit band opening at the avoided crossing of the spin-up and spin-down bands given in meV. The hopping parameters used to fit the tight-binding Hamiltonian to the DFT calculated band structure are denoted by $t_{\uparrow}$ and $t_{\downarrow}$ for spin up and spin down given in eV. Those are directional dependent for Gr$_{P.head}$ and their three values are listed. $t_R$ is the strength of the Rashba spin orbit coupling given in meV.}
\end{table*}

We discuss now the induced multiferroic-proximity effect in graphene by BFO. As we have demonstrated that the two graphene sheets exhibit different interaction strengths with the underlying BFO surface, the corresponding proximity effect is expected to differ. The calculated band structure for Gr/BFO/Gr supercell, displayed in Figure \ref{fig3} (a), reveals two distinct graphene band dispersions highlighted by blue and red corresponding to spin up and spin down, respectively. However, both graphene sheets are negatively doped which is expected due to accumulated negative charges on graphene side in response to the positive bound charges at both BFO surfaces. Following its weaker interaction with BFO, the Dirac cone corresponding to Gr$_{P.tail}$, shown in Figure \ref{fig3}(b) lies in the bulk gap of BFO closer to the Fermi level. On the other hand, the stronger interaction at Gr$_{P.head}$/BFO interface results in a larger doping of the Dirac point, as seen in Figure \ref{fig3} (c). The proximity of the insulating BFO induces modifications in the linear dispersion of the graphene band structure opening a band gap at the Dirac point. This degeneracy lifting at the Dirac point is spin dependent owing to the interaction with the magnetic BFO substrate. Interestingly, the spin-dependent band gaps and exchange splittings are influenced by the interaction strength at the BFO interface.  Spin dependent band-gaps are found to be $55$ ($26$) meV for spin up (spin down) in Gr$_{P.head}$, whereas smaller values of $6$ ($1.5$) meV are reported for Gr$_{P.tail}$. Moreover, the spin splittings for Gr$_{P.head}$ are found to be $104$ ($75$) meV for electrons (holes), respectively, compared to $35$ ($40$) meV for Gr$_{P.tail}$.      
Figure \ref{fig3} (d,e) show the evolution of the graphene band structure upon adding spin-orbit coupling to the calculations. The main impact of the spin-orbit interaction is inducing an additional band opening denoted by $\gamma_{soc}$ at the spin up/spin down band crossings. We find corresponding values of $4$ and $5$ meV for Gr$_{P.head}$ and Gr$_{P.tail}$, respectively.      

The parameters obtained from the band structure are summarized in Table~\ref{tab1} for both Gr$_{P.head}$ and Gr$_{P.tail}$. $E_{G}$ and $\Delta_{\uparrow(\downarrow)}$ represent the energy band gap and the spin dependent band gaps, respectively. The spin splitting of the electron and hole bands are denoted as $\delta_{e}$ and $\delta_{h}$. $E_{D}$ indicates how large the Dirac point doping is with respect to Fermi energy and $\gamma_{soc}$ is the spin-orbit coupling induced band opening.  The negative value of $E_{G}$ indicates a spin resolved band overlap while spin splittings are defined by spin-dependent energy differences at Dirac point with negative value indicating that spin-up bands are lower in energy than that of spin-down bands. 
Due to the stronger interaction at the {\it head} interface compared to the {\it tail}, the proximity-induced gaps and splittings are larger in Gr$_{head}$. However, the spin orbit coupling induced gap $\gamma_ {soc}$ is rather smaller. We should note here that our calculated values are different from those obtained in Ref~\cite{Qiao} due basically to the difference in the $k$-mesh size. As the band structure of graphene is highly sensitive to the k-mesh, we have used a dense $9\times 9 \times 1$ $k$-mesh in our calculations.

 The following tight-binding Hamiltonian describes the graphene's linear dispersion relation in proximity of a magnetic insulator:

\begin{equation}
\begin{split}
H = \sum_{i\sigma} \sum_l t_{l\sigma} c^\dagger_{(i+l)1\sigma} c_{i0\sigma} +h.c.
+  \sum_{i\sigma\sigma'}\sum_{\mu=0}^1 
\left[\delta+(-1)^\mu\Delta_\delta\right] c^\dagger_{i\mu\sigma} [\vec m . \vec\sigma]_{\sigma\sigma'} c_{i\mu\sigma'} \\
+  \sum_{i\sigma}\sum_{\mu=0}^1 
\left[E_D+(-1)^\mu\Delta_s\right] c^\dagger_{i\mu\sigma} c_{i\mu\sigma}, 
\end{split}
\end{equation}
where $t_{l\sigma}$ is the anisotropic hopping connecting unit cells $i$ to their nearest neighbors cells $i+l$. 
$c^\dagger_{i\mu\sigma}$ creates an electron of type ($\mu=0,1$) corresponding to A and B sites, respectively, on the unit cell $i$ with spin ($\sigma=0,1$) for spin up and spin down electrons, respectively. $\Delta_{\delta}=\frac{\delta_e - \delta_h}{2}$ where $\delta_{e}$ and $\delta_{h}$ is the strength of the exchange spin-splitting of the electron and hole bands at the Dirac cone, respectively. 
$\vec m$ is a unit vector that points in the direction of the magnetization and $\vec\sigma$ is the vector of Pauli matrices, so that  $\vec m.\vec\sigma = m_x \sigma^x+m_y\sigma^y+m_z\sigma^z$.  $E_D$ is the Dirac position with respect to the Fermi level and $\Delta_s=\frac{\Delta_\uparrow +\Delta_\downarrow}{2}$ is the averaged staggered sublattice potential.  The Rashba spin orbit coupling term is written as ~\cite{Kane,Tse}, 
\begin{equation}
H_{SO} = i t_R \sum_{i\sigma\sigma'} \sum_l  
c^\dagger_{(i+l)1\sigma} [\sigma^x_{\sigma\sigma'} d_l^x - \sigma^y_{\sigma\sigma'} d_l^y] c_{i0\sigma'} +h.c.,
\end{equation}
where $t_R$ is the Rashba spin orbit coupling strength and the vector $\vec d_l=(d_l^x,d_l^Y)$ connects the two nearest neighbors.

%\begin{equation}
%\centering
%\begin{split}
%H=\sum_{\langle i,j \rangle \alpha} t_{\alpha} c_{i \alpha} ^{\dagger}c_{j \alpha} + \frac{\delta}{2} \sum_{i \alpha} c_{i \alpha} ^{\dagger}(s_z)_{\alpha \alpha}c_{i \alpha} + \frac{\Delta_s}{2}\sum_{i \alpha} c_{i\alpha}^{\dagger}(\sigma_z)_{\alpha \alpha}c_{i \alpha}  + \\ \frac{\Delta_\delta}{2} \sum_{i \alpha} c_{i \alpha} ^{\dagger}(\sigma_z s_z)_{\alpha \alpha}c_{i \alpha} +  i\lambda_R\sum_{\langle i,j \rangle \alpha \beta}(\vec{s}_{\alpha \beta} \times d_{ij}) \cdot \hat{z} c_{i \alpha}^{\dagger}c_{j \beta}
%\end{split}
%\end{equation}
%where $c_{i \alpha} ^{\dagger}$($c_{i\alpha}$) are creation (annihilation) operators that create a particle in the site $i$ with spin $\alpha$.  $\sigma$ and $s$ are Pauli matrices that act on the sublattice and spin, respectively.  $t_{\alpha}$ is the nearest neighbours hopping parameter and it is a spin dependent term. The second term represents the exchange coupling induced in graphene, with $\delta=\frac{\delta_e+\delta_h}{2}$, where $\delta_h$ and $\delta_e$ are the strength of exchange spin-splitting of the holes and electrons, respectively. The third term represents the spin-dependent band gap opening at the Dirac point that results from the different potential felt in the sublattices A and B and $\Delta_s=\frac{\Delta_\uparrow +\Delta_\downarrow}{2}$ is the averaged staggered sublattice potential, and the fourth term represents the spin-sublattice interaction. The last term corresponds to Dirac position with respect to the Fermi level. 

To obtain the hopping values, the tight binding bands where fitted in good accordance to the DFT bands as shown by solid lines in Figures~\ref{fig3}(b-e). In the case of Gr$_{P.head}$, it was necessary to include direction dependent hopping parameters into the model. The values of the hopping parameters used for both Gr$_{P.head}$ and Gr$_{P.tail}$ are listed in Table~\ref{tab1}.

\begin{figure*}[t]
	\centering
	\includegraphics[width=1\textwidth]{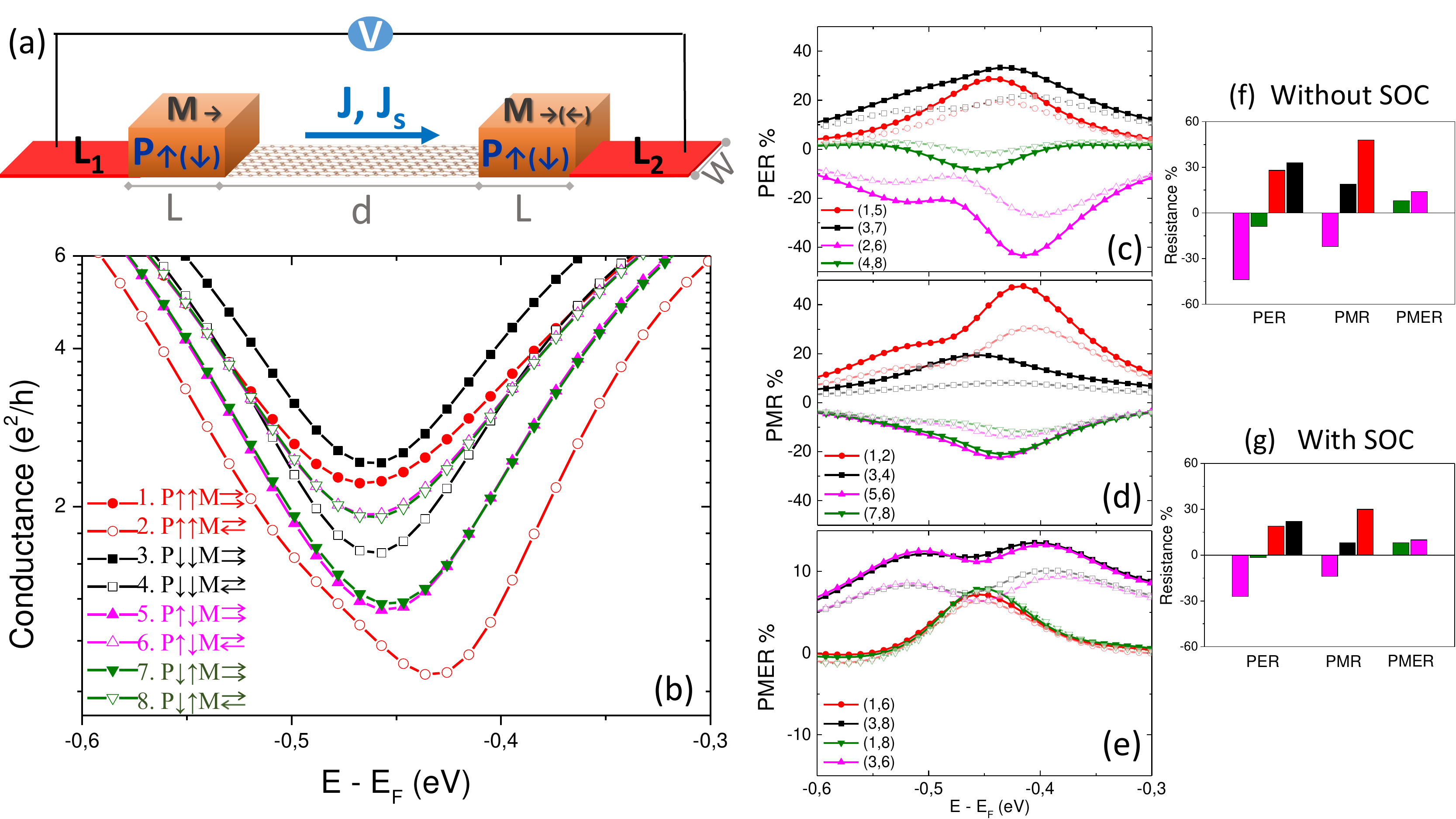}
	\caption{(color online) Model spintronic device used to calculate the multiferroic proximity magnetoresistance consisting of two multiferroic regions on top of a graphene sheet (a). The multiferroic regions have a length $L$, width $W$ and are separated by a distance $d$. (b) The conductances calculated without including spin-orbit coupling for the different configurations of electric polarization $P$ and magnetization $M$ of the two multiferroic regions. The corresponding eight conductance states are explicitly given and indexed by numbers. (c, d, e) The calculated proximity electro (PER), magneto (PMR), and multiferroic (PMER) magnetoresistances, respectively, calculated without (closed symbols) and with (open symbols) inlcuding spin orbit coupling. The indices of the two conductance states used to obtain each proximity resistance curve are designated. The maximum values of the PER, PMR, and PMER calculated without (with) including spin orbit coupling are shown in f (g), respectively.}
	\label{fig4}
\end{figure*}

Based on the Hamiltonian parameters extracted from the graphene band structure, we employed the tight-binding approach with scattering matrix formalism conveniently implemented within the KWANT package in order to calculate conductances and proximity resistances ~\cite{Kwant}. The system modeled is shown in ~\ref{fig4}(a) and comprises two identical proximity induced multiferroic regions of width $W=39.6$nm and length $L=49.2$nm, separated by a distance $d=1.5$nm of nonmagnetic region of graphene sheet with armchair edges. Both magnetic graphene regions are connected to the leads $L_{1}$ and $L_{2}$ and modeled using the Hamiltonian parameters. All the relative magnetization and polarization configurations are considered in this model device. 
The conductance $G$ in the linear response regime can be obtained according to:

\begin{equation}
G_{\alpha,\alpha'}^{\sigma,\sigma'}=\frac{e}{h}\sum_{\sigma}\int T_{\alpha,\alpha'}^{\sigma,\sigma'}\left(\frac{-\partial f}{\partial E} \right)dE ,
\end{equation}

where $T_{\alpha,\alpha'}^{\sigma,\sigma'}$ indicates spin-dependent transmission probability with ($\alpha,\alpha'$) and ($\sigma,\sigma'$) being, respectively, the relative polarization and magnetization configurations in the multiferroic regions. $f=\frac{1}{e^{(E-\mu)/k_{B}T}+1}$ is the Fermi-Dirac distribution in which $\mu$ and $T$ indicate electrochemical potential and temperature, respectively. It is important to mention that the temperature smearing has been taken into account using the room temperature since the Curie termperature of BFO is well above it. In order to show the impact of polarization on transport calculations, we choose to adjust the doping energy for the Gr$_{P.head}$ to be the same as for Gr$_{P.tail}$ bands. The conductance curves shown in Figure \ref{fig4}(b), which are explicitly described in the legend and indexed by numbers, reveal six different resistance states two of which are degenerate; those are ($5$ and $7$) and ($6$ and $8$). The conductance for a given energy should be seen as if one could gate the whole device to bring the region of interest, in the vicinity of the Dire cone splittings, to the Fermi level. We observe that the conductance curves are splitted the most in the energy range affected by proximity effect which is around $-0.47$ eV. Since the gaps and exchange splittings are much larger for Gr$_{P.head}$ compared to Gr$_{P.tail}$, a difference in the energies and conductance values between the corresponding conductance states is observed.   

The different combinations of these conductance states give rise to three types of proximity resistances:  proximity electroresistance (PER),  proximity magnetoresistance (PMR), and proximity multiferroic resistance (PMER).  
We introduce the generalized formulas of these three types of proximity resistances as follows:

\begin{equation}
PER_{\alpha,\alpha'}^{\sigma,\sigma'}=\frac{G_{\alpha,\alpha}^{\sigma,\sigma'}-G_{\alpha,-\alpha}^{\sigma,\sigma'}}{G_{\alpha,\alpha}^{\sigma,\sigma'}+G_{\alpha,-\alpha}^{\sigma,\sigma'}}
\end{equation}

\begin{equation}
PMR_{\alpha,\alpha'}^{\sigma,\sigma'}=\frac{G_{\alpha,\alpha'}^{\sigma,\sigma}-G_{\alpha,\alpha'}^{\sigma,-\sigma}}{G_{\alpha,\alpha'}^{\sigma,\sigma}+G_{\alpha,\alpha'}^{\sigma,-\sigma}}
\end{equation}

\begin{equation}
PMER_{\alpha,\alpha'}^{\sigma,\sigma'}=\frac{G_{\alpha,\alpha}^{\sigma,\sigma}-G_{\alpha,-\alpha}^{\sigma,-\sigma}}{G_{\alpha,\alpha}^{\sigma,\sigma}+G_{\alpha,-\alpha}^{\sigma,-\sigma}}.
\end{equation}

Based on this formalism, sixteen different conductance states are expected. However, due to symmetry in our considered model device we obtain $G_{\alpha,\alpha'}^{\sigma,-\sigma}=G_{\alpha,\alpha'}^{-\sigma,\sigma}$ and $G_{\alpha,-\alpha}^{\sigma,\sigma'}=G_{-\alpha,\alpha}^{\sigma,\sigma'}$ and, consequently, we end up with six conductance states $G_{\alpha,\alpha'}^{\sigma,\sigma'}$.  

The calculated PER, PMR, and PMER are plotted in Figure~\ref{fig4}(c),(d), and (e), respectively, in which the indices of the two conductance states used to obtain each proximity resistance curve are designated. Closed (open) symbol lines correspond to the calculations without (with) including SOC. Owing to the two degenerate conductance states, we obtain one (two) degenerate PMR (PMER) curves, correspondingly. The PER values range between  $-44\%$ and $+33\%$, PMR has values from $-22\%$ to $+48\%$, whereas PMER ranges between $+7\%$ and $+13\%$. We should note that including SOC doesn't change our results qualitatively but rather decreases the values of the conductances and consequently the values of the different types of proximity resistances as shown in Figure \ref {fig4}(f,g). This is basically due to the mixing of the spin channels imposed by the spin-orbit interaction. Our findings lead to a concept of multi-resistance device and pave a way towards multiferroic control of magnetic properties in two-dimensional materials. Interestingly, recent experiments have demonstrated the electric control of magnetic proximity effect at the graphene/BFO interface ~\cite{Song2} which further enhances the possibility of realizing our proposed concept device.

In conclusion, we have demonstrated that the magnetic proximity effect in graphene can be tuned by the electric polarization existing in the multiferroic substrate. The presence of electric polarization together with the polar surface charges lead to different interaction strength at the Gr/BFO interface depending on the relative direction of the electric polarization. Consequently, the spin-dependent band gaps and exchange splittings are impacted. Those findings suggest tuning the magnetic proximity effect in graphene through altering the direction or even the magnitude of the electric polarization. Such approach is accessible in multiferroic oxides where the interplay between electric and magnetic order offers the possibility of tuning the magnetization and polarization by applying electric or magnetic fields, respectively.         

We thank J. Fabian, K. Zollner  and S. Roche for fruitful discussions. This project has received funding from the European Union’s Horizon 2020 research and innovation programme under grant agreements No. 696656 and 785219 (Graphene Flagship).

\bibliography{BFO-Graphene_final}

\end{document}